\documentclass[10pt,twocolumn,showpacs,amsmath,amssymb,prl,floatfix,superscriptaddress]{revtex4-1}

\usepackage{graphicx}
\usepackage{dcolumn}
\usepackage{bm}
\usepackage{color}
\usepackage{txfonts}
\usepackage{microtype}

\begin{document}

\author{Yan-Fei Li}	\affiliation{School of Science, Xi'an Jiaotong University, Xi'an 710049, China}
\author{Rashid Shaisultanov}
\affiliation{Max-Planck-Institut f\"{u}r Kernphysik, Saupfercheckweg 1,
	69117 Heidelberg, Germany}
\author{Karen Z. Hatsagortsyan}\email{k.hatsagortsyan@mpi-hd.mpg.de}
\affiliation{Max-Planck-Institut f\"{u}r Kernphysik, Saupfercheckweg 1,
	69117 Heidelberg, Germany}
\author{Feng Wan}	\affiliation{School of Science, Xi'an Jiaotong University, Xi'an 710049, China}
\author{Christoph H. Keitel}
\affiliation{Max-Planck-Institut f\"{u}r Kernphysik, Saupfercheckweg 1,
	69117 Heidelberg, Germany}
\author{Jian-Xing Li}\email{jianxing@xjtu.edu.cn}
\affiliation{School of Science, Xi'an Jiaotong University, Xi'an 710049, China}

\bibliographystyle{apsrev4-1}

\title{Ultrarelativistic electron beam polarization in  single-shot interaction\\ with an ultraintense laser pulse}

\date{\today}

\begin{abstract}

Spin-polarization of an ultrarelativistic electron beam head-on colliding with an ultraintense laser pulse is investigated in the quantum radiation-reaction regime. We develop a Monte-Carlo method to model electron radiative spin
effects in arbitrary electromagnetic fields by employing spin-resolved
radiation probabilities in the local constant
field approximation.
Due to spin-dependent radiation reaction, the
applied elliptically polarized laser pulse polarizes the initially unpolarized electron beam and splits it
along the propagation direction into two oppositely transversely polarized parts with a splitting angle of about tens of milliradians.
Thus, a dense electron beam with above 70\% polarization can be generated in tens of femtoseconds. The proposed method demonstrates a way for relativistic electron beam  polarization with currently achievable laser facilities.

\end{abstract}

\maketitle

\emph{Introduction.} Spin-polarized electron beams have been extensively employed to investigate matter properties, atomic and molecular
structures
\cite{Kessler_1985, Getzlaff_2010, Gay_2009}.
In high-energy physics, relativistic polarized electron beams can be used to probe the nuclear structure \cite{Abe_1995,Alexakhin_2007}, generate polarized photons \cite{Maximon_1959,Martin_2012} and positrons \cite{Maximon_1959,Abbott_2016}, study  parity violation in M{\o}ller scattering \cite{Anthony_2004} and new physics beyond the Standard Model \cite{Moortgat2008}.
There are many methods to generate polarized electron beams at low energies \cite{Kessler_1985}.
However, for relativistic  electron beams, there are mainly two methods \cite{Swartz_1988}. In the first method mostly used in the Stanford Linear Accelerator,
the polarized electrons are first extracted from a photocathode (illuminated by a circularly polarized light) \cite{Pierce_1976,Pierce_2008} and then, accelerated  by the linear accelerator
(alternatively one may use polarized electrons from spin filters \cite{Batelann_1999} or beam splitters \cite{Dellweg_2017}, with subsequent laser wakefield acceleration \cite{Wen_2018}).
 The second method is a direct way of polarization of
 a relativistic electron beam in a storage ring via radiative polarization (Sokolov-Ternov effect) \cite{Sokolov_1964,Sokolov_1968,Baier_1967,Baier_1972,Derbenev_1973,Baier_1977,Derbenev_1979,Mane_1987a}. The polarization time of the latter due to the synchrotron radiation is rather slow (typically from minutes to hours), since the
magnetic fields of a synchrotron are too weak (in the order of 1 Tesla). The electrons are polarized transversely due to Sokolov-Ternov effect.
As mostly longitudinal polarization is interesting in high-energy physics,  spin rotation systems are applied \cite{Buon_1986}.
Moreover, for creating polarized positron  beams (also applicable for electrons) Compton scattering or Bremsstrahlung of circularly polarized lasers and successive pair creation are commonly used \cite{Hirose_2000,Omori_2006,Artru_2008,Muller_2011,Piazza_2010_spin}. The polarization of relativistic electrons can be detected by Compton scattering \cite{Barber_1992}, M${\o}$ller scattering \cite{Cooper_1975}, or other methods.

Strongest fields in a laboratory are provided by lasers, and the state-of-the-art ultraintense laser technology can reach a laser peak intensity in the scale of $10^{22}$ W/cm$^2$ (magnetic field strength   $\sim 4\cdot 10^5$ Tesla) \cite{Vulcan,ELI,Exawatt,Yanovsky2008}. Can such strong fields be employed to polarize electrons, similar to the Sokolov-Ternov effect? Unfortunately, previous investigations proved that electrons cannot be polarized via asymmetric spin-flip in nonlinear Compton scattering off a strong monochromatic plane laser wave \cite{Kotkin_2003,Ivanov_2004,Karlovets_2011}. In a plane-wave laser pulse the electron polarization properties due to a single photon emission
have been analyzed recently in  \cite{Seipt_2018}, and  9\% degree of polarization has been shown. It is also known that due to linear Compton scattering the electrons of different spins are scattered off the beam with different probabilities, and unscattered part of the beam becomes polarized \cite{Derbenev_1979a}, however, the number of electrons in the beam is significantly decreased in this process.
Further, recently the strong rotating electric field has been shown to highly polarize an electron beam analogous to the Sokolov-Ternov effect in tens of femtoseconds \cite{Sorbo_2017,Sorbo_2018}. The rotating electric field models anti-nodes of the electric field of a  standing laser wave. However, it is known that at available strong laser intensities the electrons are mostly trapped at nodes of the electric field, rather than anti-nodes \cite{Lehmann_2012,Kirk_2016}.

\begin{figure}[b]	
\setlength{\abovecaptionskip}{-0.2cm}  	
\includegraphics[width=1\linewidth]{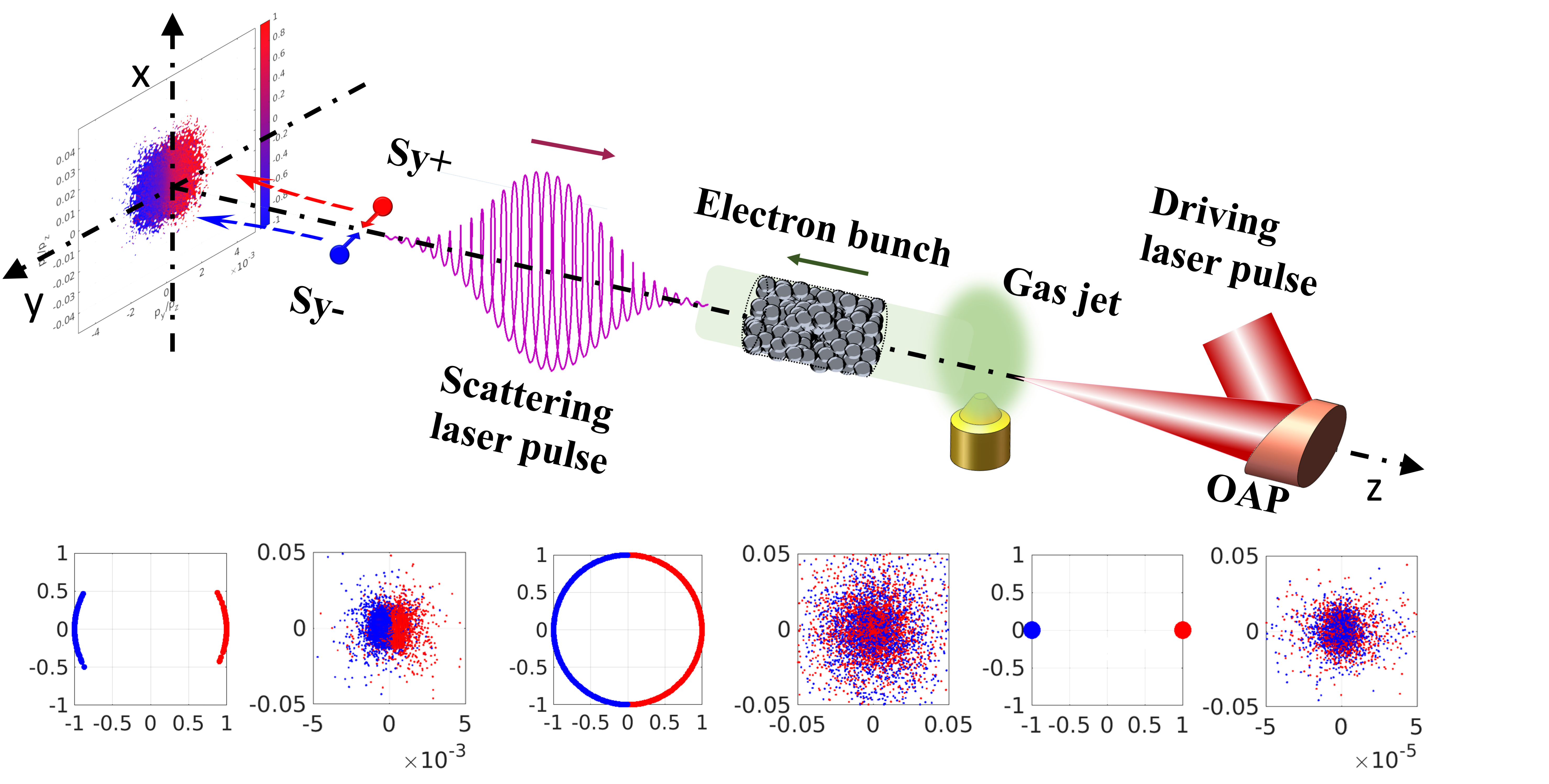}
\begin{picture}(300,0)
\put(4,125){\footnotesize (a)}
\put(20,48){\footnotesize (b)}
\put(57,48){\footnotesize (c)}
\put(95,48){\footnotesize (d)}
\put(133,48){\footnotesize (e)}
\put(171,48){\footnotesize (f)}	
\put(209,48){\footnotesize (g)}	

\put(22,11){\footnotesize  \textls[-300]{$S_y$}}
\put(-1,33){\rotatebox{90}{\footnotesize \textls[-300]{$S_x$}}}
\put(47,11){\scriptsize $p_y/p_z$}
\put(37,27){\rotatebox{90}{\scriptsize $p_x/p_z$}}

\put(98,11){\footnotesize \textls[-300]{$S_y$}}
\put(76,33){\rotatebox{90}{\footnotesize\textls[-300]{$S_x$}}}
\put(131,13){\scriptsize $p_y/p_z$}
\put(114,27){\rotatebox{90}{\scriptsize $p_x/p_z$}} 	

\put(174,11){\footnotesize \textls[-300]{$S_y$}}
\put(152,33){\rotatebox{90}{\footnotesize \textls[-300]{$S_x$}}} 	
\put(199,13){\scriptsize $p_y/p_z$}
\put(189,27){\rotatebox{90}{\scriptsize $p_x/p_z$}} 	
\end{picture}
	\caption{ Scenario of generation of spin-polarized electron beams via nonlinear Compton scattering. (a) An ultrarelativistic electron bunch generated by laser wakefield acceleration  collides head-on with an ultraintense elliptically polarized laser pulse.  ``$S_{y+}$'' (red point) and ``$S_{y-}$'' (blue point) denote the electrons polarized parallel and anti-parallel to the $y$ direction, respectively. Transverse spin distributions: for (b) EP, (d) CP, and (f) LP laser pulses. Transverse momentum distributions: for (c) EP, (e) CP, and (g) LP laser pulses. The laser pulse propagates along $+z$ direction, and the major axis of the polarization ellipse is along $x$-axis.     }
	\label{fig1}
\end{figure}
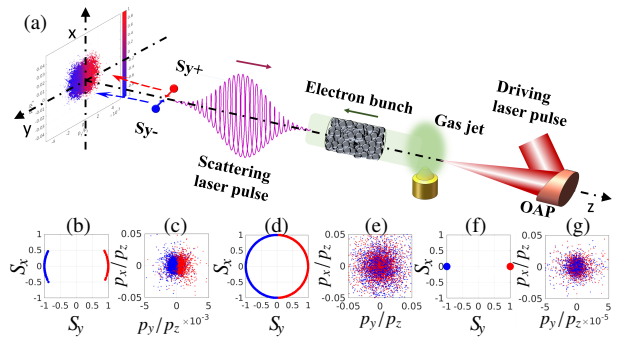

In this letter, we show that with a proper choice of ellipticity of the driving strong laser pulse, interacting with a counterpropagating unpolarized electron beam in the quantum radiation-reaction regime \cite{Piazza2012}, the electron beam can be polarized and splitted along the propagation direction into two parts, which have opposite transverse polarizations, see the interaction scenario in Fig.~1. The splitting of the electron beam  is due to  spin-dependent radiation-reaction effect.
After the interaction time of tens of femtoseconds the splitted electrons are highly polarized transversely, and the polarization rate can reach above 70\%  under currently achievable experimental conditions.
It is interesting to note that the considered effect is damped in the circularly polarized (CP) and linearly polarized (LP) laser fields, but is significant in the elliptically polarized (EP) one with a proper ellipticity, see
Figs.~\ref{fig1}(b)-(g) and the detailed explanation below in Fig.~\ref{fig3}. For the analysis of radiative spin effects we have developed a Monte-Carlo simulation method for photon emissions during the electron semiclassical dynamics in external laser field, which is based on the spin-resolved radiation probability in the local constant field approximation \cite{Piazza_2018}.

In nonlinear Compton scattering, the invariant parameter characterizing quantum effects in the strong field processes is $\chi\equiv |e|\hbar\sqrt{(F_{\mu\nu}p^{\nu})^2}/m^3c^4$ \cite{Ritus_1985}, where  $F_{\mu\nu}$ is the field tensor, $\hbar$  the reduced Planck constant, $c$ the speed of the light, $p=(\varepsilon/c,{\bf p})$  the incoming electron 4-momentum, and $-e$ and $m$ are the electron charge and mass, respectively.  When the electron counterpropagates with the laser beam, one may estimate $\chi\approx 2(\hbar\omega_0/mc^2)\xi\gamma$. Here, $\xi\equiv |e|E_0/(m\omega_0 c)$  is the invariant laser field parameter, $E_0$ and $\omega_0$ are the amplitude and frequency of the laser field, respectively, and $\gamma$ is the electron Lorentz factor.

{\it  Monte-Carlo method for the electron radiative polarization.}
We simulate in our Monte-Carlo code the space-time and the spin dynamics of electrons.
Photon emissions are treated quantum mechanically. In ultraintense laser field,  $\xi \gg1$, the coherence length of the photon emission is much smaller than the laser wavelength and the typical size of the electron trajectory \cite{Ritus_1985,Khokonov2010}. As a result, the photon emission probability is determined by the local electron trajectory, consequently, by the local value of the parameter $\chi$ \cite{Baier1994}.
The photon emission spin-dependent probabilities in the local constant field approximation are employed with the leading order contribution with respect to $1/\gamma$, which  are derived with the QED operator method of Baier-Katkov \cite{Baier_1973}:
\begin{widetext}
\begin{eqnarray}\label{Wspin}
\frac{{\rm d}W_{fi}}{{\rm d}u{\rm d}\eta}&=&W_R\left\{-(2+u)^2 \left[{\rm IntK}_{\frac{1}{3}}(u')
 -2{\rm K}_{\frac{2}{3}}(u') \right](1+{\bf S}_{if})+u^2\left[{\rm IntK}_{\frac{1}{3}}(u')
+2{\rm K}_{\frac{2}{3}}(u') \right](1-{\bf S}_{if})+2u^2({\bf S}_i\cdot{\bf S}_f){\rm IntK}_{\frac{1}{3}}(u')-\right.\nonumber\\
&&\left.(4u+2u^2)({\bf S}_f+{\bf S}_i)\left[{\bm\beta}\times\hat{{\bf a}}\right]{\rm K}_{\frac{1}{3}}(u')-2u^2({\bf S}_f-{\bf S}_i)\left[{\bm\beta}\times\hat{{\bf a}}\right]{\rm K}_{\frac{1}{3}}(u')-4u^2\left[{\rm IntK}_{\frac{1}{3}}(u')
-{\rm K}_{\frac{2}{3}}(u') \right]({\bf S}_i\cdot{\bm\beta})({\bf S}_f\cdot{\bm\beta}) \right\},
\end{eqnarray}
\end{widetext}
where $W_R={\alpha m c}/\left[{8\sqrt{3}\pi\lambdabar_c\left( k\cdot p_i\right)}{\left(1+u\right)^3}\right]$, $u'=2u/3\chi$, $u=\hbar\omega_{\gamma}/\left(\varepsilon_i-\hbar\omega_{\gamma}\right)$, ${\rm IntK}_{\frac{1}{3}}(u')\equiv \int_{u'}^{\infty} {\rm d}z {\rm K}_{\frac{1}{3}}(z)$,  ${\rm K}_n$ is the $n$-order modified Bessel function of the second kind, $\alpha$ the fine structure constant, $\lambdabar_c=\hbar/mc$ the Compton wavelength,  $\omega_{\gamma}$ the emitted photon frequency, $\varepsilon_i$ the electron energy before radiation,  $\eta=k\cdot r$ the laser phase, $p_i$, $k$, and $r$  are  4-vectors of the electron momentum  before radiation, laser wave-vector, and coordinate, respectively, ${\bm \beta}={\bf v}/c$,
$\hat{{\bf a}}={\bf a}/|{\bf a}|$ is the acceleration, 
${\bf S}_{i}$ and ${\bf S}_f$ denote the electron spin polarization vector before and after radiation, respectively, $|{\bf S}_{i,f}|=1$, and ${\bf S}_{if}\equiv {\bf S}_i\cdot{\bf S}_f$.
Summing over ${\bf S}_f$,  the radiation probability  depending on the initial spin is obtained:
\begin{eqnarray}\label{Wspin2}
\frac{{\rm d}{W}_{fi}}{{\rm d}u{\rm d}\eta}&=&8W_R\left\{-(1+u){\rm IntK}_{\frac{1}{3}}(u')
+(2+2u+u^2){\rm K}_{\frac{2}{3}}(u')\right.\nonumber\\
&&\left.-u{\bf S}_i\cdot\left[{\bm\beta}\times\hat{{\bf a}}\right]{\rm K}_{\frac{1}{3}}(u')\right\}.
\end{eqnarray}
Averaging  by the electron initial spin, the widely used radiation probability is obtained \cite{Elkina2011,Ridgers2014,Green2015, Sokolov2010}. Note that the radiation probabilities in Eqs. (1) and (2) are summed up by photon polarization.

The spin dynamics due to photon emissions are described in the spirit of the quantum jump approach \cite{Molmer_1996,Plenio_1998}, applicable when the photon formation time is much smaller than the typical time of the regular quantum dynamics. 
After a photon emission, the electron spin state is collapsed into one of its basis states defined with respect to the instantaneous spin quantization axis (SQA), which is chosen
along the magnetic field in the rest frame of electron, i.e., along ${\bm\beta}\times\hat{{\bf a}}$. We consider the stochastic spin flip at photon emission using three random numbers $N_r$, $N'_r$ and $N''_r$  in $[0, 1]$, as  follows. First, at each emission length, as the spin-dependent radiation probability in Eq.~(\ref{Wspin2}) $W_{fi}\geq N_r$, a photon is emitted. The emitted photon frequency $\omega_{\gamma}$ is determined by the condition
$\frac{1}{W_{fi}}\int_{\omega_{0}}^{\omega_{\gamma}}\frac{d W_{fi}(\omega)}{d\omega}d\omega =N'_r$. Then, the electron spin flips either parallel (spin-up) or anti-parallel (spin-down) to SQA with probabilities of $W_{fi}^{\uparrow}$ and $W_{fi}^{\downarrow}$, respectively. Here, $W_{fi}=W_{fi}^{\uparrow}+W_{fi}^{\downarrow}$, and $W_{fi}^{\uparrow}$ and $W_{fi}^{\downarrow}$ are calculated via Eq.~(\ref{Wspin}).  If $W_{fi}^{\uparrow}/{W}_{fi}\geq N''_r$, the spin flips up, otherwise, down.

Between photon emissions, the electron dynamics in the external laser field is described by Newton equations, and the spin precession is governed by the Thomas-Bargmann-Michel-Telegdi  equation
\cite{Thomas_1926,Thomas_1927,Bargmann_1959, Walser_2002}:
\begin{eqnarray}\label{spin}
\frac{{\rm d}{\bf S}}{{\rm d}\eta}&=&\frac{e\gamma}{c\left(k\cdot p\right)}{\rm S}\times\left[-\left(\frac{g}{2}-1\right)\frac{\gamma}{\gamma+1}\left({\bm \beta}\cdot{\bf B}\right){\bm \beta}\right.\nonumber\\
&&\left.+\left(\frac{g}{2}-1+\frac{1}{\gamma}\right){\bf B}-\left(\frac{g}{2}-\frac{\gamma}{\gamma+1}\right){\bm \beta}\times{\bf E}\right],
\end{eqnarray}
where ${\bf E}$ and  ${\bf B}$ are the  laser electric and magnetic fields, respectively,
$g$ is the electron gyromagnetic factor:
$g\left(\chi\right)=2+2\mu\left(\chi\right)$, $\mu\left(\chi\right)=\frac{\alpha}{\pi\chi}\int_{0}^{\infty}\frac{y}{\left(1+y\right)^3}{\bf L}_{\frac{1}{3}} \left(\frac{2y}{3\chi}\right){\rm d}y$, with ${\bf L}_{\frac{1}{3}} \left(z\right)=\int_{0}^{\infty}{\rm sin}\left[\frac{3z}{2}\left(x+\frac{x^3}{3}\right)\right]{\rm d}x$. As $\chi\ll1$, $g\approx2.00232$.
The accuracy of our Monte-Carlo code is confirmed by reproducing the well known results on the radiative polarization \cite{Sorbo_2017,Sorbo_2018,Sokolov_1968, supplemental}.

We employ a tightly-focused EP laser pulse with a Gaussian temporal profile. And, the spatial distribution of the electromagnetic fields takes into account up to $(w_0/z_r)^3$-order of the nonparaxial solution \cite{Salamin2002,Salamin2002PhysRevSTAB, supplemental}, where $w_0$ is the laser beam waist, and $z_r$ the Rayleigh length.

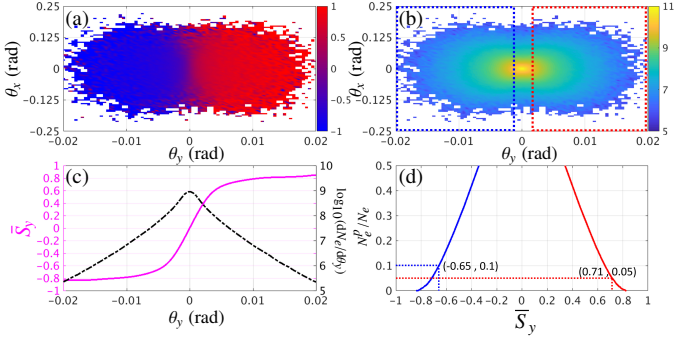
\begin{figure}[t]
	\setlength{\abovecaptionskip}{-0.4cm}
	\includegraphics[width=1.0\linewidth]{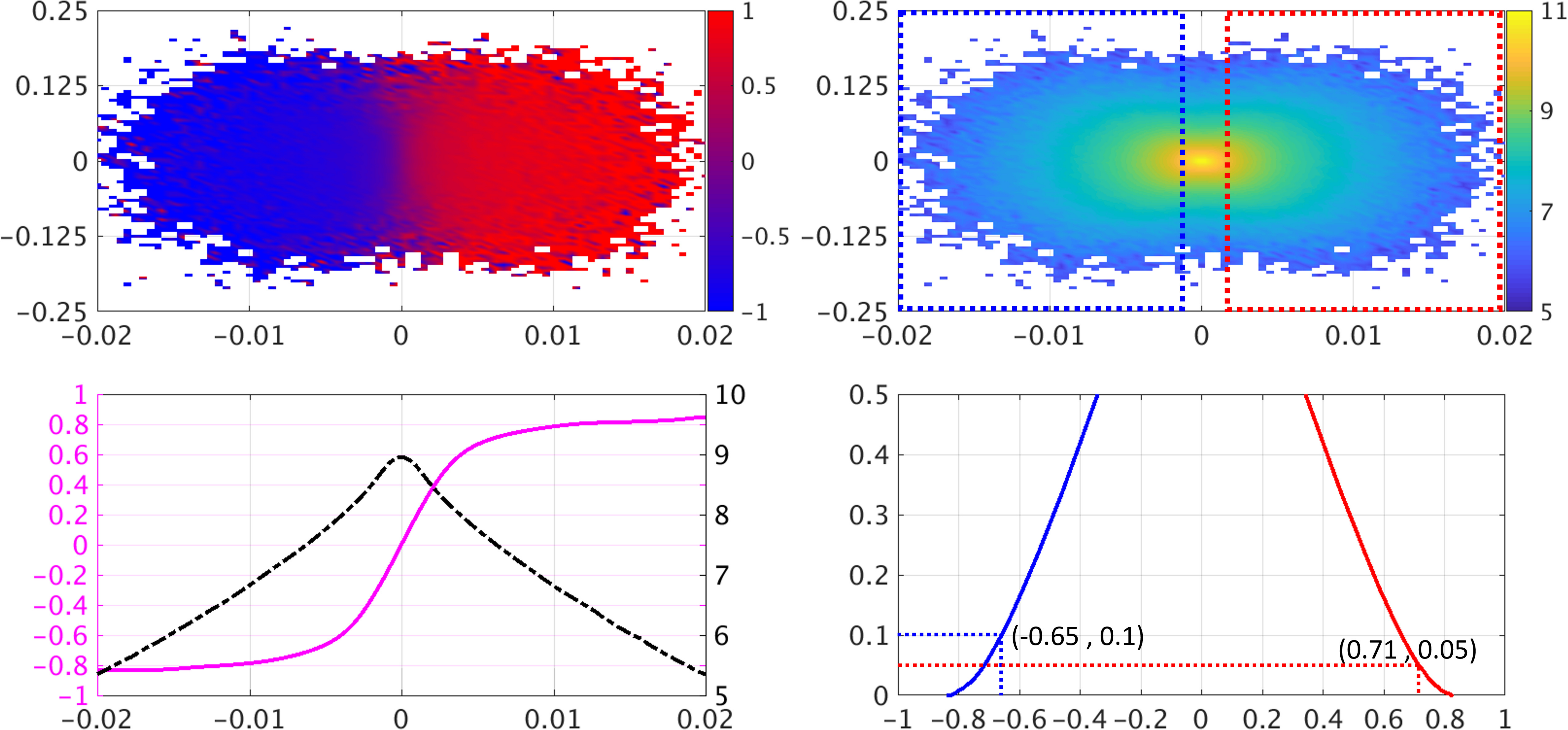}
\begin{picture}(300,25)
\put(16,132){\footnotesize (a)}
\put(142,132){\footnotesize (b)}
\put(16,72){\footnotesize (c)}
\put(142,72){\footnotesize (d)}

\put(55,81){\scriptsize $\theta_y$ (rad)}
\put(-7,104.5){\rotatebox{90}{\scriptsize $\theta_x$ (rad)}}
\put(181,81){\scriptsize $\theta_y$ (rad)}
\put(123,104.5){\rotatebox{90}{\scriptsize  $\theta_x$ (rad)}} 	
\put(55,19){\scriptsize $\theta_y$ (rad)}
\put(-5,53){\rotatebox{90}{\footnotesize \textcolor[rgb]{1,0,1}{\textls[-300]{$\overline{S}_y$}}}}
\put(186,17){\footnotesize $\overline{S}_y$}
\put(117,70){\rotatebox{-90}{\tiny   log$_{10}$(d$N_e$/d$\theta_y$)}} 	
\put(125,48){\rotatebox{90}{\tiny   $N_e^p/N_e$}}

\end{picture}
	\caption{ (a) Transverse distribution of the electron spin component $S_y$ vs the deflection angles $\theta_x={\rm arctan}(p_x/p_z)$ and $\theta_y={\rm arctan}(p_y/p_z)$; (b) Transverse distribution of the electron density log$_{10}\left({\rm d}^2 N_e/{\rm d}\theta_x{\rm d}\theta_y\right)$ rad$^{-2}$.
	(c) Average spin $\overline{S}_y$ (magenta solid) and electron distribution log$_{10}$(d$N_e$/d$\theta_y$) (black dashed) vs $\theta_y$. (d) Ratio of polarized electron number $N_e^p$ to total electron number $N_e$ vs the beam average spin $\overline{S}_y$. The red (right) and blue (left) curves represent the polarization parallel and anti-parallel to the $+y$ axis, respectively. And, the points (-0.65, 0.1) and (0.71, 0.05) indicate ($\overline{S}_y$, $N_e^p/N_e$) of electrons in the blue and red boxes in panel (b), respectively.   The laser and electron beam parameters are given in the text.  }
	\label{fig2}
\end{figure}

\textit{Results.} The considered effect of polarization of an electron beam is illustrated in Fig.~\ref{fig2}. The laser peak intensity  $I_0\approx1.37\times10^{22}$ W/cm$^2$ ($\xi=100$), wavelength $\lambda_0=1$ $\mu$m, the laser pulse duration $\tau = 5T_0$, with the laser period $T_0$, the laser focal radius $w_0=5$ $\mu$m, and the ellipticity $\epsilon=|E_y|/|E_x|=0.05$.
An electron bunch of a cylindrical form collides head-on with the laser pulse at the polar angle $\theta_e=180^{\circ}$ and the azimuthal angle $\phi_e=0^{\circ}$ with an angular divergence of 0.3 mrad. The electron initial kinetic energy $\varepsilon_0=4$ GeV ($\gamma\approx7827.8$) with an energy spread $\Delta \varepsilon_0/\varepsilon_0 =0.06$, $\chi_{max}\approx 1.5$ (the pair production is estimated to be negligible for present parameters), the electron bunch radius  $w_e= \lambda_0$, the length $L_e = 5\lambda_0$, and the density $n_e\approx 2.6\times10^{17}$ cm$^{-3}$
with a transversely Gaussian and longitudinally uniform distribution.
This kind of electron bunch can be obtained by current
laser wakefield accelerators \cite{Esarey_2009,Leemans2014}.

   \begin{figure}[t]
	\setlength{\abovecaptionskip}{-0.8cm}   	
 	\includegraphics[width=0.90\linewidth]{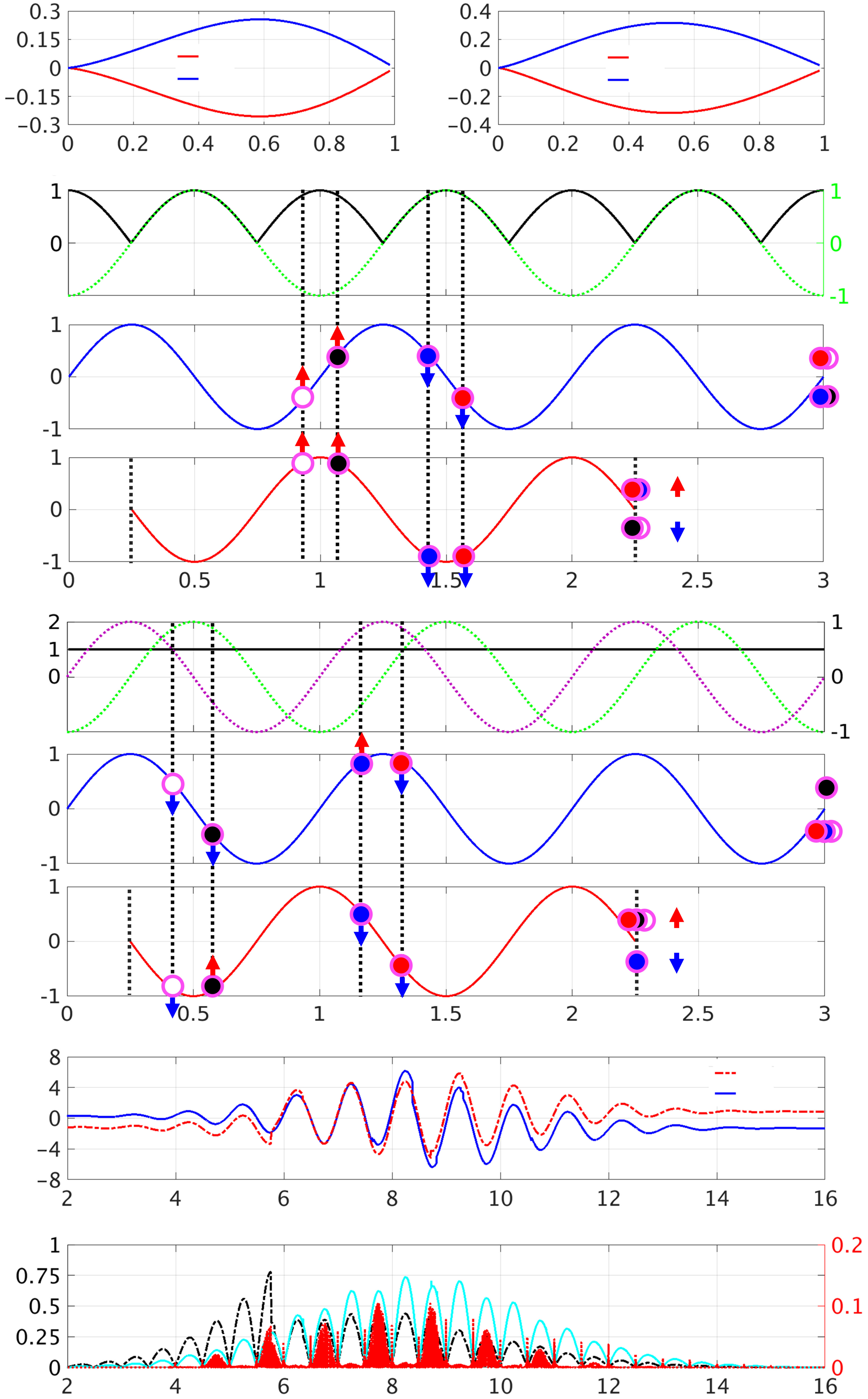}
\begin{picture}(225,32)
\put(21,381){\footnotesize (a)}
\put(56,374){\scriptsize Parallel}  	
\put(56,367){\scriptsize Antiparallel}
\put(50,345){\footnotesize $\hbar\omega_{\gamma}/\varepsilon_i$}
\put(-2,360){\rotatebox{90}{\footnotesize $\delta W_{spin}$}}

\put(131,381){\footnotesize (b)}	
\put(165,374){\scriptsize Parallel}  	
\put(165,367){\scriptsize Antiparallel}   	
\put(160,345){\footnotesize $\hbar\omega_{\gamma}/\varepsilon_i$}
\put(110,360){\rotatebox{90}{\footnotesize $\delta W_{spin}$}}

\put(0,338){\footnotesize (c)}
\put(21,316){\scriptsize (c1)}  	
\put(21,282){\scriptsize (c2)}   	
\put(110,234){\footnotesize $\eta/2\pi$}
\put(5,316){\rotatebox{90}{\scriptsize $\chi/\chi_{max}$}}
\put(5,278){\rotatebox{90}{\scriptsize $p_x/p_{x(max)}$}}
\put(5,244){\rotatebox{90}{\scriptsize $p_y/p_{y(max)}$}}
\put(21,248){\scriptsize (c3)}
\put(218,337){\rotatebox{-90}{\scriptsize {\color{green}$E_x/E_0$}}}
\put(69,286){\footnotesize $\gamma_1$}  	
\put(69,267){\footnotesize $\gamma_1$}
\put(93,295){\footnotesize $\gamma_2$}  	
\put(93,267){\footnotesize $\gamma_2$}
\put(101,295){\footnotesize $\gamma_3$}  	
\put(101,246){\footnotesize $\gamma_3$}
\put(125,286){\footnotesize $\gamma_4$}  	
\put(125,246){\footnotesize $\gamma_4$}

\put(0,227){\footnotesize (d)}
\put(21,205){\scriptsize (d1)}  	
\put(21,172){\scriptsize (d2)}
\put(21,137){\scriptsize (d3)}  	
\put(110,123){\footnotesize $\eta/2\pi$}
\put(5,205){\rotatebox{90}{\scriptsize $\chi/\chi_{max}$}}
\put(5,169){\rotatebox{90}{\scriptsize $p_x/p_{x(max)}$}}
\put(5,135){\rotatebox{90}{\scriptsize $p_y/p_{y(max)}$}}
\put(218,237){\rotatebox{-90}{\scriptsize {\color{green}$E_x/E_0$} $|$ \textcolor[rgb]{0.7,0,1}{$E_y/E_0$}}}
\put(38,188){\footnotesize $\gamma_1$}  	
\put(38,137){\footnotesize $\gamma_1$}
\put(62,175){\footnotesize $\gamma_2$}  	
\put(62,137){\footnotesize $\gamma_2$}
\put(85,192){\footnotesize $\gamma_3$}  	
\put(85,154){\footnotesize $\gamma_3$}
\put(110,192){\footnotesize $\gamma_4$}  	
\put(110,141){\footnotesize $\gamma_4$}
\put(177,153){\footnotesize Spin-up}
\put(177,142){\footnotesize Spin-down}

\put(177,263){\footnotesize Spin-up}
\put(177,252){\footnotesize Spin-down}

\put(22,112){\footnotesize (e)}    			
\put(109,76){\footnotesize $\eta/2\pi$}
\put(195,115){\scriptsize \textls[-300]{$S_y$} $+$}
\put(195,108){\scriptsize\textls[-300]{$S_y$} $-$}
\put(4,92){\rotatebox{90}{\scriptsize $p_y/mc$}}

\put(22,64){\footnotesize (f)}    			
\put(109,28){\footnotesize $\eta/2\pi$}
\put(-1,34){\rotatebox{90}{\scriptsize $\chi/2$ $|$ \textcolor[rgb]{0,1,0.9}{d$W_{rad}$/d$\eta$}}}
\put(223,64){\rotatebox{-90}{\scriptsize {\color{red} $W_{flip}$}} }

\end{picture}
	\caption{(a) and (b): The relative magnitude of the  spin-dependent term in the radiation probability of Eq.~(\ref{Wspin2})
	with $\chi=1$ and 0.1, respectively. $\delta W_{spin}\equiv W_{spin}/(W_{rad}-W_{spin})$, and, $W_{rad}$ and $W_{spin}$ are the total radiation probability and  the spin-dependent term in Eq.~(\ref{Wspin2}), respectively. Red and blue curves denote ${\bf S}_i$ parallel and anti-parallel to SQA, respectively.	(c) and (d): Electron momenta in EP (LP) and CP plane waves, respectively.
		The colored circles indicate the photon emission points in the laser field and the corresponding electron final  momenta. The red-up (blue-down) arrows indicate  ``spin-up'' (``spin-down'') with respect to $+y$ axis in (c2), (c3) and (d2), and $+x$ axis in (d3). 
		(e) Scaled $p_y$ of two sample electrons  vs $\eta$. (f) Scaled $\chi$ (black), radiation probability $W_{rad}$ (cyan) and flip probability $W_{flip}$ (red) vs $\eta$ for a sample electron.   The laser and electron beam parameters in (e) and (f) are the same as in Fig.~\ref{fig2}. }
	\label{fig3}
\end{figure}
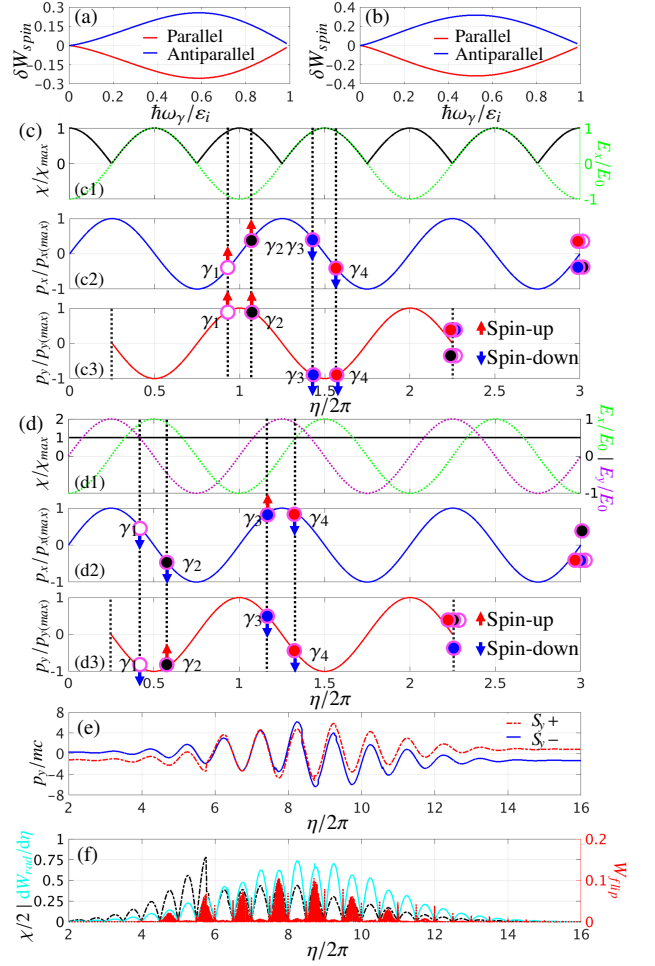

The simulation results presented in Fig.~\ref{fig2} show that an initially unpolarized electron bunch is polarized and splitted into two beams polarizing parallel and anti-parallel to the minor axis of elliptical polarization ($+y$ axis), respectively, with a splitting angle of about 20 mrad, see Fig.~\ref{fig2}(a), which is much larger than the angular divergence of the electron beams \cite{supplemental}.  The corresponding electron density mainly concentrates in the beam center, since  the transverse ponderomotive force
is relatively small, see Fig.~\ref{fig2}(b). Figure~\ref{fig2}(c) represents the average spin $\overline{S}_y$ (magenta-solid curve) and the electron density distribution (black-dashed curve) integrated over $\theta_x$.
Near $\theta_y=0$, the electron density is rather high, but $\overline{S}_y$ is very low. With the increase of $|\theta_y|$, the electron density exponentially declines, however, $\overline{S}_y$ remarkably ascends until about 80\%. Separating the part of the electron beam within $\theta_y>0$ (or $\theta_y<0$), one will obtain an electron beam with positive (or negative) transverse polarization.
 When splitting the beams exactly at $\theta_y=0$, one obtains $|\overline{S}_y|\approx 34.21\%$  for both of splitted beams. However, we can increase the polarization of beams if we exclude the electrons near $\theta_y=0$. For instance,  as is shown by blue and red boxes in Fig.~\ref{fig2}(b), the corresponding average spin $\overline{S}_y$ and electron number ratio $N_e^p/N_e$ are approximately (-65\%, 10\%) and (71\%, 5\%), respectively, see Fig.~\ref{fig2}(d). The corresponding splitting angle is of about 3 mrad, which is much larger than the angular resolution (less than 0.1 mrad) with current technique of electron detectors \cite{Wang2013,Leemans2014,Wolter2016,Chatelain2014}.

Moreover, for experimental convenience, we consider the cases of larger energy spread $\Delta \varepsilon_0/\varepsilon_0=0.1$, larger angular divergence of 1 mrad and different collision angles $\theta_e=179^\circ$ and $\phi_e=90^\circ$, and all show stable and uniform results \cite{supplemental}.

The reason for the electron beam polarization and splitting is analyzed in Fig.~\ref{fig3}. The spin effect in the radiation probability is due to the third term in Eq.~(\ref{Wspin2}), and
its contribution is rather significant (about 30\%)
for high-energy photon emission, see Figs.~\ref{fig3}(a), (b) at $\hbar\omega_{\gamma}/\varepsilon_i \approx  0.5\sim0.6$, which is
 negative (positive) for ${\bf S}_i$ parallel (anti-parallel) to SQA.
For simplicity, we analyze the electron radiative dynamics in plane wave cases, see Figs.~\ref{fig3}(c), (d).  Let us assume that the relativistic electrons initially move along $-z$ direction, have no transverse momentum, and the final polarization along  $y$ axis is detected.
When in the laser field the electron emits a photon (mostly at large $\chi$)
 with a transverse momentum, finally it will appear with an opposite  one due to the momentum conservation. The ultrarelativistic electron is assumed to emit a photon along its momentum direction, since the emission angle $\sim 1/\gamma$ is rather small. Therefore, the electron final transverse momentum will be opposite by sign to its momentum at the photon emission point.
In the laser field the transverse momentum ${\bf p}_\bot=e{\bf A}(\eta)$, with the vector potential ${\bf A}(\eta)$,  is delayed by $\pi/2$ with respect to the field ${\bf E}(\eta)$.
The SQA is along ${\bm\beta}\times\hat{{\bf a}}\propto e{\bm\beta}\times{\bf E}+e{\bm\beta}\times({\bm \beta}\times{\bf B})\sim e(1-{\bm\beta}_z){\bm\beta}\times{\bf E}$, and note that ${\bm\beta}$ is negative.

For the LP plane wave polarized along $x$ axis, see Figs.~\ref{fig3}(c1), (c2), $\chi\propto\xi\gamma$ oscillates with $|E_x|$, and the SQA is along $y$ axis, with a sign following $E_x$.
According to  Eq.~(\ref{Wspin2}) and  Figs.~\ref{fig3}(a), (b), at points of $\gamma_1$, $\gamma_2$, the photon emission is more probable for spin-up (with respect to $+y$ direction) electrons, because the corresponding $E_x$ (green curve), and consequently, SQA are both negative. At points of $\gamma_3$ and $\gamma_4$, spin-down electrons mostly radiate.
The final transverse momenta of electrons emitting photons at $\gamma_1$ and $\gamma_4$ are positive and at $\gamma_2$ and $\gamma_3$ negative. Consequently, spin-up and spin-down electrons move symmetrically with respect to the $x$ axis and mix together, as indicated in Figs.~\ref{fig1}(f), (g).

For the CP plane wave, see Figs.~\ref{fig3}(d1)-(d3), $\chi$ is constant, and the SQA rotates along the propagation $z$ axis. In Fig.~\ref{fig3}(d2), at points of $\gamma_1$ and $\gamma_2$, spin-down (with respect to $+y$ direction) electrons more probably radiate (since the corresponding $E_x$ and $y$ component of SQA are both positive), and final $p_x<0$ for $\gamma_1$ and $p_x>0$ for $\gamma_2$. The similar analysis applies for other points, e.g., for $\gamma_3$ the final $p_x<0$ (spin-up) and for $\gamma_4$, $p_x<0$ (spin-down). Thus, spin-up and spin-down electrons mix together with respect to $x$ axis.
 Similar electron spin dynamics exist for $p_y$ in Fig.~\ref{fig3}(d3) as well. Finally,
spin-up and spin-down electrons mix together in $x-y$ plane, as indicated in Figs.~\ref{fig1}(d), (e).

 However, for the EP plane wave with a rather small ellipticity ($E_y\ll E_x$), the radiation probability and the SQA both mainly relies on $E_x$, and the SQA is along $y$ axis. In Fig.~\ref{fig3}(c3), $p_y$ has a $\pi$ delay with respect to $E_x$; at points of $\gamma_1$ and $\gamma_2$, $E_x$ and SQA are both negative, thus, spin-up (with respect to $+y$ direction) electrons more probably radiate and finally acquire negative $p_y$. And, at points of $\gamma_3$ and $\gamma_4$, spin-down electrons more probably radiate and finally have positive $p_y$. Consequently, electrons split up with respect to the $+y$ axis, see Figs.~\ref{fig1}(b),(c) and \ref{fig2}, in which, since $p_z$ is negative, spin-up (spin-down) electrons move at positive (negative) $\theta_y=$ arctan$(p_y/p_z)$. The trajectories of sample electrons in Fig.~\ref{fig3}(e)  illustrate those behaviors.

We underline that the considered effect of the spin-dependent splitting of the beam relies  on the spin-dependent radiation reaction, rather than on the asymmetric spin flip. Moreover, multiple flips of spin will smear out the considered effect and we judiciously have chosen parameters to reduce the flip effect  via limiting the number of emitted photons: $N_{ph}\sim \xi \alpha\tau/T_0\approx3.65$ \cite{Piazza2012,supplemental}, along with rather small spin flip probability, see Fig.~\ref{fig3}(f).

 \begin{figure}[t]
	\setlength{\abovecaptionskip}{-0.1cm}
	\includegraphics[width=1.0\linewidth]{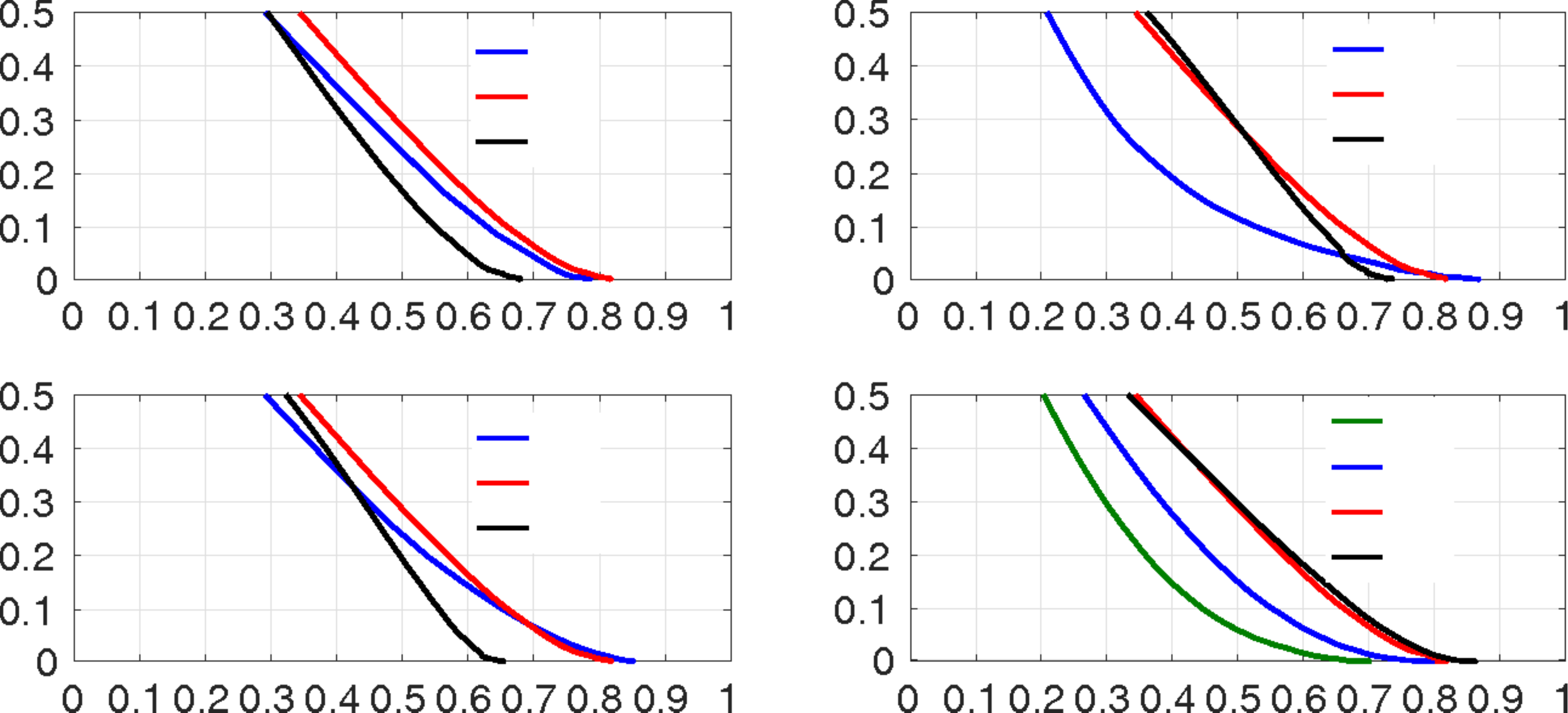}
\begin{picture}(0,0)(100,0)
\put(-10,113){\footnotesize (a)}
\put(121,113){\footnotesize (b)}
\put(-10,53){\footnotesize (c)}
\put(121,53){\footnotesize (d)}

\put(-32,88){\rotatebox{90}{\scriptsize $N_e^p/N_e$}}
\put(38,63){\scriptsize \textls[-300]{$\overline{S}_y$}}
\put(64,112.5){\scriptsize $\epsilon \rm{=}0.01$}
\put(64,105){\scriptsize $\epsilon \rm{=}0.05$}
\put(64,97.5){\scriptsize $\epsilon \rm{=}0.1$}

\put(98,88){\rotatebox{90}{\scriptsize $N_e^p/N_e$}}
\put(170,63){\scriptsize \textls[-300]{$\overline{S}_y$}}
\put(198,113){\scriptsize $\xi=50$}
\put(198,105.5){\scriptsize $\xi=100$}
\put(198,97.5){\scriptsize $\xi=150$}

\put(-32,27){\rotatebox{90}{\scriptsize $N_e^p/N_e$}}
\put(38,2){\scriptsize \textls[-300]{$\overline{S}_y$}}
\put(64,52){\scriptsize $\tau=3T_0$}
\put(64,44){\scriptsize $\tau=5T_0$}
\put(64,36.5){\scriptsize $\tau=10T_0$}

\put(98,27){\rotatebox{90}{\scriptsize $N_e^p/N_e$}}
\put(170,2){\scriptsize \textls[-300]{$\overline{S}_y$}}
\put(197,54){\scriptsize 0.5 GeV}
\put(200,47){\scriptsize  1  GeV}
\put(200,39.5){\scriptsize 4  GeV}
\put(199,32.5){\scriptsize 20 GeV}
\end{picture}
	\caption{Impacts of (a) ellipticity $\epsilon$, (b) laser intensity $\xi$, (c) laser pulse duration $\tau$, and (d) initial kinetic energy of electrons $\varepsilon_0$ on the polarization. Other parameters are the same as in Fig.~\ref{fig2}. }
	\label{fig4}
\end{figure}
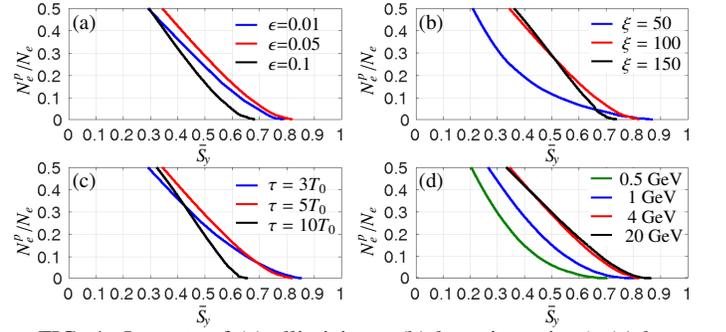

 Furthermore, impacts of the laser and electron beam parameters on the polarization are analyzed in Fig.~\ref{fig4}. First, the ellipticity $\epsilon$ is a very crucial parameter. If $\epsilon$ is too small, the splitting angle $\theta_s\sim p_{\perp}/p_{\parallel}\propto E_y/E_x$ is very small as well, and the polarized electrons partially overlap near $p_y=0$ (e.g., the ultimate case of the LP laser), which reduces the degree of polarization.
 Oppositely,
 largely increasing ellipticity can increase the splitting angle, but unfortunately also the SQA rotation (cf., the ultimate case of the CP laser). As a result the average polarization decreases, see Fig.~\ref{fig4}(a). The  optimal  ellipticity is of order of $10^{-2}$ to $10^{-1}$. The trade off exists also for the laser intensity, pulse duration, and the electron energy. From one side, the effect relies on the radiation reaction and requires large $\chi\approx 10^{-6}\xi\gamma\gtrsim 1$ and many photon emission. From another side, the spin flips smear out the considered effect which imposes restriction on the photon emissions. For this reason,
 with increasing $\xi$ and the electron kinetic energy $\varepsilon_0$, the polarization is first enhanced due to the increase of $\chi$, and then saturates, see Fig~\ref{fig4}(b), (d).
 The mentioned trade off yields nonuniform dependence on the laser pulse duration. The polarization is weak at too short or too long pulses, and the optimum is $\tau=5T_0$ for the given parameters, see Fig~\ref{fig4}(c).

For a simple  estimation of radiative polarization effects, we also develop a semi-classical  analytical method based on the modified Landau-Lifshitz equation \cite{Landau1975, Poder2018, Piazza2012} with a radiation-reaction force accounting for quantum-recoil and  spin effects.
This model further confirms above obtained results qualitatively \cite{supplemental}.

In conclusion, we have developed a  Monte-Carlo method for simulating radiative spin effects. We show that adding a proper small ellipticity to the strong laser pulse allows to directly polarize and split a counterpropagating relativistic electron beam into highly polarized parts with current achievable experimental techniques, which can be used in high-energy physics.

 {\it Acknowledgement:}  We are grateful to A. Di Piazza and M. Tamburini for helpful discussions. This work is supported by the  Science Challenge Project of China (No. TZ2016099), the National Key Research and Development Program of China (Grant No. 2018YFA0404801), and the National Natural Science Foundation of China (Grants Nos. 11874295, 11804269).

\bibliography{QEDspin}

\end{document}